# Heterogeneous length of stay of hosts' movements and spatial epidemic spread

Chiara Poletto[1], Michele Tizzoni[1,§], Vittoria Colizza[2,3,4,*]

[1] Computational Epidemiology Laboratory, Institute for Scientific Interchange (ISI), Torino, Italy
[2] INSERM, U707, Paris, France
[3] UPMC Université Paris 06, Faculté de Médecine Pierre et Marie Curie, UMR S 707, Paris, France
[4] Institute for Scientific Interchange (ISI), Torino, Italy

[§] Current address: Mathematical Modelling Services, UBC Centre for Disease Control, 655 West 12th Ave., Vancouver BC V5Z 4R4, Canada.

[*] Corresponding author : vittoria.colizza@inserm.fr

**Infectious diseases outbreaks are often characterized by a spatial component induced by hosts' distribution, mobility, and interactions. Spatial models that incorporate hosts' movements are being used to describe these processes, to investigate the conditions for propagation, and to predict the spatial spread. Several assumptions are being considered to model hosts' movements, ranging from permanent movements to daily commuting, where the time spent at destination is either infinite or assumes a homogeneous fixed value, respectively. Prompted by empirical evidence, here we introduce a general metapopulation approach to model the disease dynamics in a spatially structured population where the mobility process is characterized by a heterogeneous length of stay. We show that large fluctuations of the length of stay, as observed in reality, can have a significant impact on the threshold conditions for the global epidemic invasion, thus altering model predictions based on simple assumptions, and displaying important public health implications.**

Space represents, in many circumstances, a relevant feature in the spread of an infectious disease[1]. This is the case of recent outbreaks that reached the global scale, such as e.g. the SARS outbreak in 2002-2003[2] and the 2009 H1N1 pandemic influenza[3,4,5], but also applies to country or region-specific epidemics[6] or even outbreaks at smaller scales[7,8]. The diffusion of a directly transmitted disease in a given environment depends on the spatial distribution of susceptible hosts and their ability to move from one region to the other of the space, connecting otherwise isolated communities.

The recent availability of high-resolution large datasets of hosts' population spatial distribution and mobility have enabled the development of spatially based modelling approaches that, by integrating the above knowledge with the disease description, are able to provide a theoretical and quantitative framework to describe the disease behaviour in time and space[1,9-14]. Hosts' movements represent a key ingredient of such approaches, and different modelling approximations are made depending on the type of mobility process, or based on simplification considerations, or due to the constraints imposed by limited available knowledge. Movements may be permanent, as in the case of human migrations[15] or livestock trade displacements[16,17], or may be characterized by origin-destination patterns with subsequent return to the origin, as in the case of human travel in general. The latter case requires the explicit introduction of an element of memory in the process, as well as the specification of the duration of the hosts' visit at destination. Short trips within Europe during holiday period in spring-summer 2009, for example, contributed to the diffusion of the 2009 H1N1 pandemic in the continent, following the first European outbreaks seeded by Mexico and the US[18]. Different travel habits and trip durations, induced by geographical, cultural and seasonal conditions, may have different outcomes. In particular, the timescale of the length of stay at destination is an important aspect of the spatial spreading process, as it represents the time window during which the disease can be brought to the unaffected population at destination by an infectious traveller visiting that given destination, or during which the healthy traveller may contract the disease from the outbreak taking place at destination while visiting (as e.g. in the H1N1 case above). Given its relevance for spreading processes, and the large variability of individual travel behaviour affecting the reasons for travel and associated frequency and durations of trips, here we consider a modelling approach that explicitly includes this timescale and its fluctuations, aiming at the understanding of the inclusion of additional ingredients in ever more realistic data-driven models. Several degrees of heterogeneities are indeed found to characterize many aspects relevant to epidemic spreading – from the individual level[19-22] to the population level [10,16,23,24] – and to have a profound impact on epidemic phenomena occurring on such complex substrates[22,25-31]. Here, through a metapopulation theoretical framework and detailed simulations, we address the inclusion of an extra degree of complexity, namely the heterogeneous length of stay of hosts at destination, and discuss its impact with respect to simpler modelling assumptions.

## Results

**Metapopulation model with heterogeneous length of stay**

We consider a population of hosts that is spatially structured into $V$ subpopulations coupled by hosts' movements, representing a metapopulation network where nodes correspond to subpopulations where the infection dynamics takes place, and links correspond to the mobility between origin and destination subpopulations. Analyses of human mobility data and transportation infrastructures have shown the presence of various levels of heterogeneities characterizing the movement patterns[8,31-35]. The structure of these patterns often displays large variability in the number of connections per mobility centre (whose corresponding geographic census area represents the subpopulation in the metapopulation framework), and broad fluctuations are also observed in the traffic exchanged by each mobility centre or flowing on a given connection between origin and destination. In Figure 1 we report the example of the European air transportation network[36], showing the probability distributions of the number of connections $k$ per airport (i.e. the subpopulation's degree, panel a) and of the total traffic of passengers $w_{ij}$ travelling between any pair of linked airports *i* and *j* (Figure 1b). In addition, general statistical laws can be found that relate the traffic flow between any two geographic areas and their properties, such as population sizes, number of connections or other features[9,10,23,34,37], depending on the specific system under study. In the case of the air transportation network, the behaviour of the average weight $w_{kk'}$ flowing along the connection between two airports with degree $k$ and $k'$ is found to be a function of the product of the degrees[26], i.e. $\langle w_{kk'}\rangle \propto (kk')^\theta$. Empirical data on the population sizes of the geographical census areas associated to mobility patterns also show scaling relations in terms of the subpopulation degree $k$, with $N_k \propto k^\phi$ where $N_k$ is the population size of the census area with $k$ air-travel connections[9]. These features have been observed at various resolution scales and for different types of mobility modes[9,26,37,38,39], including commuting processes[38] and within-cities daily movements[27,39].

In order to capture these empirical properties, we consider a metapopulation model characterized by a random connectivity pattern described by an arbitrary degree distribution $P(k)$, and in the following we will explore the role of realistic heterogeneous network structures and also homogeneous network structures for comparison. To take into account the large degree fluctuations empirically observed, we describe the system by adopting a degree-block notation[25] that assumes statistical equivalence for subpopulations of equal degree $k$[27]. This corresponds to assuming that all subpopulations having the same number of connections are considered statistically identical. In other words, subpopulations with degree $k$ are characterized by the same population size $N_k$ and by the same mobility properties, including the length of stay and the traffic of individuals, as illustrated below. While disregarding more specific properties of each individual subpopulation – that may be related for instance to local, geographical or cultural aspects – this mean field approximation is able to capture the degree dependence of many system's properties as observed in the data, and also allows for an analytical treatment of the system's behavior[27].

Following the scaling properties observed in real-world mobility data, we define: (i) the population size $N_k$ of a subpopulation with degree $k$ as $N_k = \bar{N} k^\phi / \langle k^\phi \rangle$, with $\bar{N} = \sum_k N_k P(k)$ being the average subpopulation size of the metapopulation system; (ii) the number of individuals moving from the subpopulation of degree $k$ to the subpopulation of degree $k'$ as $w_{kk'} = w_0 (kk')^\theta$. The exponents $\phi$ and $\theta$, and the scaling factor $w_0$ assume different values depending on the application to the mobility process under consideration. The mobility of hosts along the links of the metapopulation network is modelled with the per capita diffusion rate $\sigma_{kk'} = w_{kk'} / N_k$ where we assume that individuals are randomly chosen in the population according to the patterns $w_{kk'}$. The overall leaving rate out of the subpopulation with degree $k$ is given by $\sigma_k = k \sum_{k'} \sigma_{kk'} P(k'|k)$ where $P(k'|k)$ is the conditional probability that a node with degree $k$ is connected to a node with degree $k'$, and we define the diffusion rate rescaling as $\sigma = w_0 \langle k^\phi \rangle / \bar{N}$. Finally, individuals return to their origin after an average time $\tau_{k'}$ that corresponds to the length of stay at destination with degree $k'$, or the duration of their visit.

The economic literature considers the length of stay as one of the key elements that need to be solved in modelling a visitor's decision-making process[38]. Its determinants, however, are still largely debated and no clear consensus has emerged on the problem[39]. Empirical data on the length of stay at the individual level is rather scant, but many statistical sources clearly indicate a vast heterogeneity characterizing this quantity. Figure 1c reports the probability distribution of the length of stay of people travelling to and from the United Kingdom for all purposes, showing large fluctuations ranging from 1 day or less, up to several months[40]. Similar broad distributions are observed in the travel patterns of visitors that spend their holidays in Europe (Figure 1d), and from other sources of travel statistics as reported in the Supplementary Information (SI). Next to air travel, heterogeneous duration of visits are also observed in human trajectories characterizing daily activities[8,32,34], ranging from few minutes to several hours. To incorporate these fluctuations in the metapopulation modelling framework, we assume that the length of stay at destination $\tau_{k'}$ is a scaling function of the degree $k'$ of the subpopulation of destination, i.e.

$$\tau_{k'} = \frac{\bar{\tau}}{\langle k^\chi \rangle} k'^\chi \qquad (1)$$

where $\bar{\tau}/\langle k^\chi \rangle$ is the normalization factor, and $\bar{\tau} = \sum_k \tau_k P(k)$ is the average length of stay on the metapopulation network. By tuning the value of the power-law exponent $\chi$, we can define different regimes of the mobility dynamics. For $\chi > 0$ the length of stay is positively correlated with the degree of the subpopulation of destination, meaning that individuals travelling to a well-connected location will spend a longer time at destination, thus being longer exposed to the local population, with respect to individuals travelling to peripheral locations. This can be motivated by the attractiveness of popular locations, for which the pattern of connection is optimized through large degrees to manage large fluxes volumes of individuals, both at the within-city scale and at the larger geographical scale where airport hubs handle large traffic due to tourism or seasonal/temporary job opportunities[41,42]. The opposite regime, obtained for $\chi < 0$, implies that the time spent at a location is larger for decreasing degree of the subpopulation of destination, and may correspond to an individual choice of optimization between the time spent to reach the destination and the time spent at destination. Low degree locations are indeed generally peripheral in the system, so that a trip from a given origin may take multiple steps to reach the final destination, and thus a longer length of stay at the hard-to-reach destination may then adequately balance and justify the travel time[43]. The value $\chi = 0$ corresponds to the case of homogeneous length of stay as it is generally assumed in the recurrent mobility process of commuting where the time $\tau$ represents the duration of an average working day[10,16,30], whereas the condition $\chi \to \infty$ corresponds to the case of permanent migration that is used to model mobility processes with no return to the origin (such as the case of livestock displacements in trade flows[16,17]) or to approximate origin-destination mobility by simplifying the modelling approach and assuming a Markovian process[27,44]. Here, instead, we are interested in understanding whether if and to what extent the empirically observed heterogeneity of the length of stay may affect the invasion dynamics of the disease and its geotemporal spread, by assuming that this heterogeneity is fully encoded in the connectivity properties of the subpopulation of destination. The assumption on the geographical dependence of the length of stay, determined by the degree $k$ of the location, finds its support in travel statistics available at the city level that aggregate various travel modes and reasons (see the SI). Empirical evidence from higher detail datasets on human movements that would allow us to identify a specific functional form for Eq. (1) is currently lacking, therefore we adopt a rather simple power-law behaviour, consistently with the other scaling properties of human mobility, enabling the analytical study of the epidemic invasion process in the system under variations of the assumed parameters while preserving the heterogeneity of $\tau$. More sophisticated assumptions can be made on the expression of $\tau$, that may depend, for instance, both on the origin and destination subpopulations, or on the individual behaviour, and will be the object of future studies.

In order to take into account the memory effects associated with the mobility process (i.e. the return to the location of origin), we subdivide the population of individuals $N_k$ resident in the subpopulation of degree $k$ into those individuals who are from $k$ and are located in $k$ at time $t$, $N_{kk}(t)$, and those who are from $k$ and are located in a neighbouring subpopulation having degree $k'$ at time $t$, $N_{kk'}(t)$. A schematic representation of the system, illustrating the subdivision of the populations for two connected locations, is shown in Figure 2, and the definition of the degree-block variables is reported in Table 1. This formalism[10,30,45,46] allows keeping track of the origin subpopulation, and the mobility process is mathematically represented by rate equations describing the time behaviour of $N_{kk'}(t)$ in terms of the diffusion and return rates, as illustrated in the Methods section. The overall dynamics is characterized by the interplay between the timescales of the mobility process and the intrinsic timescale of the infectious disease. Realistic values of the mobility rates fall in the regime $\sigma_k \ll \tau_k^{-1}$, as the fraction of travellers of a population on the typical timescale of the movement is very small, as for instance observed from air travel[9] and commuting data[10]. By focusing on infectious diseases characterized by relatively long timescales, a quasi-stationary approximation can be considered that assumes the mobility process to be at equilibrium with respect to the epidemic process[10,30,46], with the subpopulation sizes assuming their stationary values:

$$N_{kk} = \frac{\bar{N}}{\langle k^\phi \rangle} v_k k^\phi \tag{2}$$

$$N_{kk'} = \frac{\sigma \bar{N} \bar{\tau}}{\langle k^\phi \rangle \langle k^\chi \rangle} v_k k^\theta k'^{\theta+\chi} \tag{3}$$

with the factor $v_k$ given by $v_k = \left(1 + \sigma\bar{\tau}\frac{\langle k^{\theta+\chi+1}\rangle}{\langle k\rangle\langle k^\chi\rangle}k^{\theta-\phi+1}\right)^{-1}$ (see the Methods section for the full derivation).

While mobility connects different locations, the epidemic process is modelled within each subpopulation, where we assume a standard SIR model that partitions the total number of individuals into susceptible, infectious, and recovered individuals[47]. The SIR infection dynamics is ruled by the transition rates, $\beta$ and $\mu$, representing the disease transmissibility rate (for the transition from susceptible to infectious) and the recovery rate (for the transition from infectious to recovered), respectively. The epidemic model is characterized by the reproductive number[47] $R_0 = \beta/\mu$, that defines the average number of infected individuals generated by one infectious individual in a fully susceptible population, thus leading to the threshold condition $R_0 > 1$ for an epidemic in a single population[47].

The quasi-stationary approximation adopted to describe the subpopulation sizes in terms of Eqs. (2) and (3) is ensured by considering an infectious period longer than the typical length of stay at destination, i.e. $\mu^{-1} \gg \tau_k$. This is well supported in the case of many relevant examples, such as influenza or SARS with $\mu^{-1}$ of the order of days, when the metapopulation framework is applied to the geotemporal scale of human daily activities, characterized by broad distributions of time spent at each activity location[8,32,34] however limited by a cutoff of approximately 17 hours capturing the typical awake interval of an individual[8]. In the following we will solve the system in the relatively long infectious period approximation: we will characterize the dependence of the invasion behaviour on the length of stay parameters $\chi$ and we will show through numerical simulations that the behaviour is still valid when the approximation does not hold anymore, making this approach applicable to a variety of geotemporal scales regarding infectious diseases and types of mobility, ranging from hourly activities to daily and monthly trips.

**Threshold condition for global invasion**

While $R_0$ provides a threshold condition for the occurrence of the outbreak, the epidemic spread at the global level is clearly dependent on the spatial structure of the population. An outbreak may indeed start in a seeded subpopulation, but it may not be able to spread to the neighbouring subpopulations due to specific conditions related to the mobility process. For example, the diffusion rate may be small enough as to prevent the travel of an infectious individual before the epidemic ends in the seeded subpopulation; or, the time spent at destination by the travelling infectious individual is too short for her/him to transmit the disease to individuals of the local population before returning to the subpopulation of origin. The occurrence of these events has a clear stochastic nature and is captured by the definition of an additional predictor of the disease dynamics, $R_*$, regulating the number of subpopulations that become infected from a single initially infected subpopulation[47-50], analogously to the reproductive number $R_0$ at the individual level. In the following, we show that it is possible to provide an analytical expression for this threshold parameter assuming non-homogeneous origin-destination patterns of mobility captured by Eq. (1), leading to non-trivial effects in the spreading dynamics, and going beyond the cases of permanent migration[28,33] and homogeneous mobility processes[30].

Let us consider the invasion process of the epidemic spread at the metapopulation level, by using the subpopulations as our elements of the description of the system, in a Levins-type modelling approach. We assume that the outbreak starts in a single initially infected subpopulation of a given degree $k$ and describe the spread from one subpopulation to the neighbour subpopulations through a branching process approximation[51]. We denote by $D_k^n$ the number of infected subpopulations of degree $k$ at generation $n$, with $D_k^0$ being the initially seeded subpopulation, $D_k^1$ the subpopulations of degree $k'$ of generation 1 directly infected by $D_k^0$ through the mobility process, and so on. By iterating the seeding events, it is possible to describe the time behaviour of the number $D_k^n$ of infected subpopulations as follows:

$$D_k^n = \sum_{k'} D_{k'}^{n-1}(k'-1)P(k|k')\left(1 - R_0^{-\lambda_{k'k}}\right)\left(1 - \sum_{m=0}^{n-1}\frac{D_k^m}{V_k}\right) \qquad (4)$$

The r.h.s. of equation (4) describes the contribution of the subpopulations $D_{k'}^{n-1}$ of degree $k'$ at generation $n-1$ to the infection of subpopulations of degree $k$ at generation $n$. Each of the $D_{k'}^{n-1}$ subpopulations has $k'-1$ possible connections along which the infection can spread. The infection from $D_{k'}^{n-1}$ to $D_k^n$ occurs if: (i) the connections departing from nodes with degree $k'$ point to subpopulations with degree $k$, as ensured by the conditional probability $P(k|k')$; (ii) the reached subpopulations are not yet infected, as indicated by the probability $\left(1 - \sum_{m=0}^{n-1}\frac{D_k^m}{V}\right)$, where $V_k$ is the number of subpopulations with degree $k$; (iii) the outbreak seeded by $\lambda_{k'k}$ infectious individuals travelling from subpopulation $k'$ to subpopulation $k$ takes place, and the probability of such event is given by $\left(1 - R_0^{-\lambda_{k'k}}\right)$[52]. The latter term is the one that relates the dynamics of the local infection at the individual level to the coarse-grained view that describes the disease invasion at the metapopulation level. Given the subpopulation of degree $k'$ where an outbreak is taking place, the spread of the infection to a neighbouring disease-free subpopulation of degree $k$ can occur due to the travel of infected individuals resident in $k'$ who interact with the population of $k$ during their visit, or to the infected individuals resident in $k$ who come back to their subpopulation of origin after a trip to the subpopulation $k'$. If we denote by $\alpha$ the attack rate of the SIR epidemic, i.e. the total fraction of the population that is infected by the epidemic, then we can express the number of seeds as[30] $\lambda_{k'k} = \alpha(N_{kk'} + N_{k'k})$ where we assume the stationary expressions for the populations $N_{kk'}$ and $N_{k'k}$ given by Eqs. (2) and (3). By plugging this expression into equation (4), and assuming an epidemic unfolding on an uncorrelated metapopulation network with a reproductive number close to the threshold value, $R_0 \simeq 1$, it is possible to analytically solve equation (4) for the early stage of the epidemic process, and obtain the following expression for the global invasion threshold:

$$R_* = \frac{2(R_0-1)^2}{R_0^2} \sigma \overline{N} \overline{\tau} \frac{\Lambda(\{P(k)\}, \sigma, \overline{\tau}, \overline{N})}{\langle k \rangle \langle k^\phi \rangle \langle k^\chi \rangle} \tag{5}$$

where

$$\Lambda(\{P(k)\}, \sigma, \overline{\tau}, \overline{N}) = \langle (k-1)k^{2\theta+\chi+1} v_k \rangle + \sqrt{\langle (k-1)k^{2(\theta+\chi)+1} \rangle \langle (k-1)k^{2\theta+1} v_k^2 \rangle} \tag{6}$$

is a functional form of the degree distribution and of its moments, of the average length of stay $\overline{\tau}$, of the leaving rate rescaling $\sigma$, and of the average subpopulation size $\overline{N}$ (see the Methods section for the full derivation).

Equation (5) defines the threshold condition for the global invasion: if $R_*$ assumes values larger than 1, the epidemic starting from a given subpopulation will reach global proportion affecting a finite fraction of the subpopulations; if instead $R_* < 1$, the epidemic will be contained at its source without invading the metapopulation system. While assuming a simple SIR model with homogeneous mixing for the disease dynamics, it is important to note that stochasticity and discreteness of the mobility events are fundamental aspects to be considered in order to obtain the expression for $R_*$. A continuum approximation would indeed lead to invasion under all conditions of mobility and length of stay, after the initial outbreak in the seed, due to the movements of anyhow small fractions $\sigma_{kk'}I_k$ of infectious individuals[48-50,53].

**Impact of length of stay**

The global invasion threshold $R_*$ is a combination of several ingredients of the system, including disease parameters, demography, metapopulation network structure, travel fluxes and mobility timescales. Figure 3 shows the role played by the topological heterogeneity of the metapopulation network and the topologically coupled heterogeneous length of stay, by reporting the condition $R_* > 1$ in the comparison between different network structures (panel a). A heterogeneous metapopulation network structure with size $V = 10^4$ and characterized by a power-law degree distribution $P(k) \propto k^{-\gamma}$ with $\gamma = 3$ and average degree $\langle k \rangle = 3$ is compared to a homogeneous random topology of the same size, where every node is assumed to have a degree equal to $\langle k \rangle$. When no heterogeneity in the length of stay is considered, i.e. for $\chi = 0$, heterogeneous structures dramatically favour the epidemic invasion, as indicated by the threshold $R_* = 1$ occurring at smaller values of the reproductive number with respect to the homogeneous one. This means that, in this regime, there is an interval in the $R_0 > 1$ region for which the disease transmission potential is small enough so that the epidemic can be contained in a homogeneous system, but large enough to reach global proportion on a population that is heterogeneously spatially structured. This finding confirms previous results on the role of large degree fluctuations in reducing the threshold value for the disease invasion in a metapopulation system with Markovian[28,33] or recurrent homogeneous mobility[30]. When $\chi \neq 0$, this effect is complicated by the inclusion of an additional layer of complexity. By increasing the value of $\chi$, $R_*$ considerably increases its value, leading to a corresponding reduction of the critical value of the reproductive number above which the global invasion occurs, thus further favouring the epidemic spread even for very small $R_0$ by means of an additional mechanism. This is obtained by keeping fixed the average value of the length of stay $\bar{\tau}$. Travelling hosts spend a longer time visiting largely connected subpopulations than peripheral ones, further enhancing the spreading potential of the hubs[27] and thus making the disease propagation on the metapopulation system increasingly more efficient. On the contrary, negative and decreasing values of $\chi$ tend to suppress the critical spreading potential of the hubs, by balancing their large degree with increasingly smaller lengths of stay. The degree heterogeneity is counterbalanced by the fluctuations of the length of stay expressed by its negative correlations with the subpopulation degree; the result is the increase of the critical threshold in the value of $R_0$ that distinguishes containment from disease invasion. This counterbalancing effect is not observed in the case of a homogeneous network, due to the absence of degree fluctuations. In the regime $\chi < 0$, the spreading potential of the heterogeneous metapopulation network induced by the degree fluctuations is increasingly lowered by the chosen approximation for the length of stay, until it effectively reduces to the homogenous case, as displayed in Figure 3. In some circumstances, for given values of the parameters, the spreading potential in the heterogeneous case becomes smaller than in the corresponding homogeneous one, due to the suppressing role of hubs characterized by very small length of stay, as reported in the SI.

In the interval [-1,1] of $\chi$ under consideration, the absolute variation on the critical value of $R_0$ corresponding to the invasion threshold condition is approximately equal to 30%, and its relative variation on the outbreak condition $R_0 > 1$ is of more than 90% (Figure 3a), thus showing the importance of the fluctuations in the length of stay in the outbreak management. Smaller or larger variations can be obtained as they depend on the specific metapopulation system under study, the type of mobility, the structure of the network and the geotemporal resolution scale considered. For instance, ground workflows are found to be one order of magnitude larger than air travel flows[10], thus affecting the value of $\sigma$ in Eqs. (5) and (6), whereas different levels of degree heterogeneity can be found depending on the type of mobility network or the region under study. In the Supplementary Information we provide additional examples of the invasion region $R_* > 1$ by varying the model parameters that are application-specific.

Next to the invasion threshold condition $R_* = 1$ it is also important to investigate the absolute value of the predictor $R_*$ above the threshold, as its distance from the critical value, i.e. $R_* - 1$, is a quantitative measure of the public health efforts that need to be put in place for the containment of the disease. Figure 3b shows the 3D surface of the global invasion threshold $R_*$ as a function of the reproductive number $R_0$ and of the parameter $\chi$ tuning the heterogeneity of $\tau$. The values of $R_*$ rapidly increase in the invasion region, reaching very high values even for set of $(R_0, \chi)$ values close to the critical ones. While maintaining fixed the mobility network structure and other system features (such as e.g. the population size of each subpopulation), it is possible to envision control strategies in terms of reductions in the leaving rate occurring during the outbreak, thus encoded in reductions of the leaving rate rescaling $\sigma$ in Eq. (5). These would correspond to the application of travel-related intervention measures but also to the effects of self-reaction of the population who avoid travelling to the affected area, as observed during the early stage of the SARS outbreak and of the 2009 H1N1 pandemic[54]. The magnitude of such reduction would likely need to be, however, extremely large to bring the value of $R_*$ below the threshold in a range of realistic parameter values, as observed from the 3D surface, thus explaining the failure of feasible travel restrictions aimed at the containment of the disease[54].

**Synthetic networks and numerical simulations**

In order to test the validity of our analytical treatment, we performed extensive Monte Carlo numerical simulations at the individual level, by keeping track of the disease dynamics, of the movement and of its associated memory for each host in the system. Simulations are based on discrete-time stochastic processes, as detailed in the Methods section. We explored the phase space in order to make a direct comparison with the analytical findings discussed above, and considered heterogeneous and homogeneous metapopulation networks having the same size and average degree. More in detail, consistently with the uncorrelated approximation used in the analytical framework developed, we consider heterogeneous uncorrelated random networks generated through the uncorrelated configuration model[55] that allows the creation of a network structure characterized by a given degree distribution $P(k)$ and having no topological correlations. Homogeneous random networks are obtained by generating Erdős Rényi graphs[56] through wiring each couple of nodes with probability $\langle k \rangle/(V-1)$, were $\langle k \rangle$ is the prescribed average node degree, and degrees are distributed according to a Poisson distribution. Once the networks are constructed, all other demographical and mobility quantities are also defined, based on the scaling relations assumed for the length of stay $\tau_k$ (Eq. (1)), the population $N_k$, the diffusion rate $\sigma_{kk'}$, where different diffusion scenarios can be explored by tuning the diffusion rate rescaling $\sigma$. By fixing the average length of stay $\bar{\tau}$, we explore values of the exponent $\chi$ in the interval [-1, 0.4] to ensure the applicability of the time-scale separation approximation as well as feasible computational times. Starting from one seeded subpopulation, we simulate 500 stochastic realizations of the spatial epidemic simulation averaging on different initial conditions, and on different realizations of the metapopulation network that constitute the spatial structure of the population under study. For each simulation we can calculate the number of infectious individuals in time in the whole system and in each subpopulation, and evaluate the number of subpopulations reached by the outbreak as a function of time.

Figure 4 shows the final size of the epidemic, expressed in terms of the simulated final fraction of infected subpopulations in the system, $D_\infty/V$, as a function of $\chi$ and $R_0$, for the heterogeneous network topology (panels a and b) and for the comparison between the heterogeneous and the homogeneous case (panels c and d, respectively). The threshold behaviour is shown through the transition from the containment region (where $D_\infty/V = 0$) to the invasion region of the parameter space, where a finite fraction of the subpopulations of the system is reached by the infection. The role of the diffusion rate is shown in Figure 4a where different transitions are obtained as a function of $\chi$ for the same value of the reproductive number. As expected from equation (5), at fixed values of $\chi$ and $R_0$, larger diffusion rates favor the spread and can bring the system above the threshold. The dependence of the invasion behaviour on $\chi$ and $R_0$ is reported in Figure 4b. For smaller values of $\chi$, the critical threshold in $R_0$ increases; the smaller the value of $\chi$ and the larger needs to be the transmission potential of the disease so that it can reach global proportion and spread throughout the metapopulation system. For $\chi = 0.4$, instead, only diseases very close to the epidemic threshold $R_0 = 1$ are contained at the source. Given the role of the diffusion rate in the invasion process, as shown in Figure 4a, we tested that the enhancement of the spreading at the system level due to an increase of the value of $\chi$ is not related to an increase of the volume of people travelling, induced by the constraint of fixing the average length of stay across all numerical experiments. By imposing a constant $\bar{\tau}$, the increase of $\chi$ corresponds actually to a slight decrease of the average total traffic per link, i.e. the sum of people leaving and coming back to a given subpopulation in the quasi-stationary approximation, as illustrated in the Methods section. The global invasion is therefore favoured by a more efficient spreading mechanism allowed by the presence of central nodes characterized by a large number of connections and large visiting times. Finally, we report on the good agreement between the individual-level simulations and the corresponding analytical results presented in the previous subsection, as shown by the vertical and color-coded arrows in the Figure indicating the analytical values of the transition.

The effect of the heterogeneity of the metapopulation network structure on the global invasion threshold is further tested in Figure 4c and d where the simulation results obtained by comparing two different network topologies, heterogeneous vs. homogeneous structures with equal size and average degree, respectively, are shown, all other demographic, mobility, and disease parameters ($R_0 = 1.2$) kept equal. Global invasion is reached for a wide range of $\chi$ and $\sigma$ values explored in the heterogeneous case, whereas the invasion occurs only for a considerably smaller set of values of $\chi$ and $\sigma$ in the case of a homogeneous metapopulation network.

Next to the transition behaviour, the effect of degree-correlated heterogeneous length of stay is also evident in the spatial propagation of the epidemic in the system above the invasion threshold. By focusing, in this analysis, on simulated epidemics starting from the same initial conditions, it is possible to build the epidemic invasion tree that represents the most probable transmission of the infection from one subpopulation to the other during the history of the epidemic[10]. The stochastic nature of the epidemic process implies that each realization will produce a different tree. An overall epidemic invasion network can be constructed by weighting each link of propagation from subpopulation $i$ to subpopulation $j$ with the probability of occurrence of the transmission along the mobility connection $i \to j$ over the various stochastic realizations, and then the corresponding minimum spanning tree can be extracted to eliminate loops and focus on the main directed paths of transmission[10] (see the Methods section for further details).

Figure 5 presents a visualization of the invasion trees obtained for a heterogeneous network where positively and negatively-correlated lengths of stay are considered. A node is chosen at random and it is used as the starting condition of 100 stochastic realizations of the SIR epidemic in both cases. The resulting trees can be mapped out in terms of successive layers of infection from the origin, with nodes in the first layer ordered by degree (size of the dot) and by seeding time (color), showing how different values of $\chi$ alter the hierarchy of the epidemic invasion from one subpopulation to another. For $\chi > 0$, largely connected subpopulations have a predominant role in the further spatial spread of the disease, thanks to the two-fold favouring property of having a large degree and a longer visiting time. A different picture is obtained for $\chi < 0$, where the spreading potential of hub subpopulations, due to their high degree, is suppressed by the very short time duration that individuals spend there. More peripheral subpopulations become instead mainly responsible of the spreading dynamics towards the rest of the system.

**Role of the timescale assumption**

The results presented in the above subsections are obtained in the relatively long infectious period assumption, i.e. $\mu^{-1} \gg \tau_k$, to ensure the applicability of the timescale separation technique and allow the comparison between analytical and numerical findings. In the simulations we assumed an infectious period $\mu^{-1} = 500$ time steps expressed in arbitrary time units. The validity of such approximation clearly depends on the specific application and the spatial and temporal resolution considered. If we focus on human daily activities, based e.g. on the data obtained from cellular phones[34,35] or high-resolution surveys[7,8,24], realistic values of the mobility timescale of the system range from few minutes to few hours, translating the infectious period $\mu^{-1} = 500$ time steps in a duration of approximately few days, assuming the minimum timestep of the system to be equal to few minutes. The developed framework thus allows the study of infections such as influenza-like-illness spreading on a metapopulation network of locations visited during daily activities. A metapopulation model applied to the air travel mobility, often used for the study of the global spread of infectious diseases[5,10,44], would instead offer a different picture. Here the typical timescale of the system is of the order of days, thus an infectious period of $\mu^{-1} = 500$ timesteps, as chosen in the simulations, would correspond to 500 days given the minimum timestep of the system being equal to 1 day, and thus it would be much larger than the value corresponding to infectious diseases like influenza or SARS. In order to test the validity of our assumption on the length of the infectious period, we perform additional simulations by considering SIR-like diseases characterized by increasingly smaller $\mu^{-1}$ values, and evaluate the effect of the heterogeneous length of stay on the invasion dynamics.

Figure 6 shows the results of the transition observed from the containment to the invasion region expressed in terms of the final fraction $D_\infty/V$ of infected subpopulations, as a function of the reproductive number $R_0$ (panel a) and of the parameter $\chi$ (panel b), and for different SIR-like diseases. We explored values of the infectious period that range from $\mu^{-1} = 500$ to $\mu^{-1} = 4$ timesteps, by keeping the same average length of stay $\bar{\tau}$ as in the previous results. The cases $\mu^{-1} = 4$ and $\mu^{-1} = 10$, once expressed in days, correspond to the length of the infectious period of diseases as influenza and SARS, respectively, and therefore they allow us to explore the validity of the framework applied to a metapopulation model coupled by air travel mobility, where the timestep of the simulation is typically 1 day[5,10,44].

While a transition indicating the presence of an invasion threshold is recovered, the quantitative values of the threshold show a dependence on $\mu$ that is not found in the analytical expression of Eq. (5), obtained under the relatively long infectious period approximation. By changing $\mu^{-1}$ of two orders of magnitude, however, the relative variation observed in the $R_0$ threshold value is quite limited and approximately equal to 10%. In addition, the dependence of the global invasion threshold on the parameter $\chi$ tuning the length of stay is maintained, as reported in Figure 6b for different values of the diffusion rates, and $\mu = 0.05$. While the analytical predictions are affected by the break-down of the timescale approximation, numerical results show the robustness of the threshold behaviour of the metapopulation system, indicating the applicability of the introduced theoretical framework to a wide range of geotemporal scales, and the importance of heterogeneous lengths of stay in the perspective of epidemic control.

Finally, we report in Figure 7 on the spatial invasion pattern by studying the epidemic invasion trees in the case of $\mu^{-1} = 20$, corresponding to the SIR metapopulation dynamics of Figure 6b with $\sigma = 10^{-4}$. Even in the absence of relatively long infectious periods, the resulting pattern show the critical role of $\chi$ in shaping the hierarchy of the invasion and the spreading role of subpopulations in different degree classes.

## Discussion

Heterogeneities have long been recognized to be important in the spreading of an infectious disease, both at the contact level between individuals and at the connection level between subpopulations of individuals. Here, we have considered an extra layer of heterogeneity characterizing the host dynamics in terms of the duration of visits and found how large fluctuations in this quantity strongly alter the conditions for the disease invasion. Given the assumed dependence of the length of stay with the subpopulation degree, both the theoretical framework and the applied numerical simulations have shown that two regimes are found that may dramatically favour or hinder the invasion, induced by the positive or negative degree-correlation, respectively, altering the predictions of simpler models. Despite its simplicity, the present framework uncovered an important aspect that must be considered in interpreting and simulating epidemic spreading patterns, and in providing detailed model predictions. As the spatial spread plays a crucial role in the management and control of a disease, the present results call for the need for higher resolution mobility data to better characterize the length of stay and include additional realistic aspects – such as, e.g., the heterogeneity of the travel behaviour of the population, the dependence of mobility rates on the distance travelled, and others – that may enable an increasingly accurate description of the disease propagation.

# Methods

**Mobility at the equilibrium**

The rate equations describing the non-Markovian travelling dynamics (see SI for more details) can be written in terms of the variables $N_{kk}(t)$ and $N_{kk'}(t)$ by adopting a degree-block notation that assumes statistical equivalence for subpopulations of equal degree $k$[27]

$$\partial_t N_{kk}(t) = -\sigma_k N_{kk}(t) + k \sum_{k'} N_{kk'}(t) P(k'|k) \tau_{k'}^{-1}$$
$$\partial_t N_{kk'}(t) = \sigma_{kk'} N_{kk}(t) - N_{kk'}(t) \tau_{k'}^{-1}$$

where $\sigma_k = k \sum_{k'} \sigma_{kk'} P(k'|k)$ represents the overall leaving rate out of $k$, and the conditional probability $P(k'|k)$ represents the probability that a node with degree $k$ is connected to a node with degree $k'$. The condition $\partial_t N_{kk}(t) = \partial_t N_{kk'}(t) = 0$ yields the equilibrium solutions, equations (2) and (3). We first write the equilibrium relation $N_{kk'} = N_{kk} \sigma_{kk'} \tau_{k'}$, and then plug it into the expression for the total number of individuals resident in the subpopulation $k$, $N_k = N_{kk} + k \sum_{k'} N_{kk'} P(k'|k)$. By making explicit the variable $N_{kk}$ in the latter equation and considering that for uncorrelated networks the conditional probability is given by the expression $P(k'|k) = kP(k)/\langle k \rangle$, we recover the expressions (2) and (3). The full derivation is reported in the SI.

From equations (2) and (3) we can compute $T_{kk'}$, the total volume of people travelling along each link at the equilibrium, that is the sum of people resident in $k$ and travelling to their destination $k'$, and people resident in $k'$ returning after visiting $k$

$$T_{kk'} = \frac{\sigma \bar{N}}{\langle k^\phi \rangle} (kk')^\theta (\nu_k + \nu_{k'}).$$

$T_{kk'}$ is function of all the parameters of the system, and in particular of the exponent $\chi$. Therefore, scenarios with different values of $\chi$ differ in the value of the average total traffic, since we impose the same average length of stay $\bar{\tau}$. For the case of uncorrelated network, the average total traffic volume $\langle T_{kk'} \rangle$ is given by the expression

$$\langle T_{kk'} \rangle = \sum_{kk'} \frac{kP(k) k' P(k')}{\langle k \rangle^2} T_{kk'} = \frac{2\sigma \bar{N}}{\langle k \rangle^2 \langle k^\phi \rangle} \langle k^{\theta+1} \nu_k(\chi) \rangle \langle k^{\theta+1} \rangle,$$

where we have highlighted the dependency on $\chi$. $\langle T_{kk'} \rangle$ decreases as the value of $\chi$ increases with a relative variation no greater than 2% considering the two cases $\chi = -1$ and $\chi = 0.4$ (all other parameter values set as in the main text).

**Calculation of the global invasion threshold**

In order to obtain the explicit form for equation (4) we plug into the equation the expression for the number of seeds $\lambda_{kk'} = \alpha(N_{kk'} + N_{k'k})$, where $N_{kk'}$ and $N_{k'k}$ are given by the equilibrium relation (3). For the attack rate $\alpha$ we take the approximate expression valid under the condition[57] $R_0 \cong 1$, $\alpha \cong \frac{2(R_0-1)}{R_0^2}$. We obtain in such a way the following expression for the equation (4)

$$D_k^n = C\left[k^{\theta+1}v_k P(k)\sum_{k'} D_{k'}^{n-1}(k'-1)k'^{\theta+\chi} + k^{\theta+\chi+1}P(k)\sum_{k'} D_{k'}^{n-1}(k'-1)k'^{\theta}v_{k'}\right]$$
(7)

where $C = \frac{2(R_0-1)^2}{R_0^2}\frac{\sigma\bar{N}\bar{\tau}}{\langle k\rangle\langle k^\phi\rangle\langle k\chi\rangle}$. At this point we can write a close form of the above iterative process by defining the vector[30] $\boldsymbol{\Theta}^n = (\Theta_1^n, \Theta_2^n)$, where

$$\Theta_1^n = \sum_k (k-1)k^{\theta+\chi}D_k^n$$
$$\Theta_2^n = \sum_k (k-1)k^\theta v_k D_k^n$$

The dynamics equations of $\boldsymbol{\Theta}^n$ are encoded in equation (7) and can be written as $\boldsymbol{\Theta}^n = C\mathbf{G}\boldsymbol{\Theta}^{n-1}$, with $\mathbf{G}$ being the two-dimensional matrix with elements

$$g_{11} = g_{22} = \langle(k-1)k^{2\theta+\chi+1}v_k\rangle$$
$$g_{12} = \langle(k-1)k^{2(\theta+\chi)+1}\rangle$$
$$g_{21} = \langle(k-1)k^{2\theta+1}v_k^2\rangle$$

The dynamics is driven by the largest eigenvalue of $\mathbf{G}$, which defines the function

$$\Lambda(\{P(k)\}, \sigma, \bar{\tau}, \bar{N}) = \langle(k-1)k^{2\theta+\chi+1}v_k\rangle + \sqrt{\langle(k-1)k^{2(\theta+\chi)+1}\rangle\langle(k-1)k^{2\theta+1}v_k^2\rangle}$$

appearing in the equation (5) of the global invasion threshold parameter.

**Numerical simulations**

We explicitly simulate both the travelling among different subpopulations and the infection transmission within each subpopulation as discrete-time stochastic processes, treating individuals as integer units. At each time step, travelling individuals along with new infectious and recovering individuals for each subpopulation are extracted randomly from binomial and multinomial distributions. More in detail, the system is updated according to the following rules. (a) Infection dynamics: (i) The contagion process assumes the homogenous mixing within each population, specifically at each time step the force of infection $F_k$ within a subpopulation of degree $k$, is given by $F_k = \beta\frac{I_k^*}{N_k^*}$, where $I_k^*$ and $N_k^*$ are respectively the total number of infectious and the total number of individuals present at that moment in the subpopulation, i.e. $I_k^* = I_{kk} + k\sum_{k'} I_{k'k}P(k'|k)$, and $N_k^* = N_{kk} + k\sum_{k'} N_{k'k}P(k'|k)$; (ii) At the same time, each infectious individual is subject to the recovery process with rate $\mu$. (b) After all nodes have been updated for the local infection process, we simulate the diffusion process by randomly extracting for all nodes the number of individuals departing to each of the $k$ destination and the ones coming back.

**Epidemic invasion tree**

The epidemic invasion tree is a directed, weighted minimum spanning tree among all the possible propagation paths starting from the same initial condition, extracted as follows. For each subpopulation pair $lj$, we define $p_{lj}$ as the probability of infection transmission from $l$ to $j$. This probability shows the likelihood that subpopulation $j$'s infection is seeded by subpopulation $l$. This can happen by two means: either a resident in $j$ acquires the infection in $l$ and brings it home, or an infectious traveller from $l$ brings the infection to $j$. Then, $p_{lj}$ is defined as the proportion of runs, where $j$ has been seeded by $l$. Finally, we define a distance metric $d_{ij} = \sqrt{1 - p_{lj}}$ to measure dissimilarities for the infection probability. The minimum spanning tree is then calculated using the Chu-Liu-Edmunds Algorithm[58].

## References

1. Riley, S. Large-scale transmission models of infectious disease. *Science* **316**, 1298-1301 (2007).
2. McLean, A. R., May, R. M., Pattison, J. & Weiss, R. A. *SARS. A Case Study in Emerging Infections* (Oxford University Press; 2005).
3. Fraser, C., *et al.* Pandemic potential of a strain of influenza A/H1N1: early findings. *Science* **324**, 1557-1561 (2009).
4. Khan, K., *et al.* Spread of a novel influenza A(H1N1) virus via global airline transportation. *N. Engl. J. Med.* **361,** 212-214 (2009).
5. Balcan, D., *et al.* Seasonal transmission potential and activity peaks of the new influenza A(H1N1): a Monte Carlo likelihood analysis based on human mobility. *BMC Med.* **7**: 45 (2009).
6. Keeling, M. J. Models of foot-and-mouth disease. *Proc. R. Soc. B* **272**, 1195-1202 (2005).
7. Eubank, S., *et al.* Modelling disease outbreaks in realistic urban social networks. *Nature* **429**,180-184 (2004).
8. Eubank S, Kumar VSA, Marathe MV, Srinivasan A, Wang N. Structure of social contact networks and their impact on epidemics. AMS-DIMACS Special Volume on Epidemiology (2006).
9. Colizza, V., Barrat, A., Barthélemy, M. & Vespignani, A. The role of the airline transportation network in the prediction and predictability of global epidemics. *Proc. Natl. Acad. Sci. USA* **103**, 2015-2020 (2006).
10. Balcan, D., Colizza, V., Gonçalves, B., Hu, H., Ramasco J. J. & Vespignani, A. Multiscale mobility networks and the large scale spreading of infectious diseases, *Proc. Natl. Acad. Sci. USA* **106**, 21484-21489 (2009).

11. Ferguson, N. M., *et al.* Strategies for containing an emerging influenza pandemic in Southeast Asia. *Nature* **437**, 209 – 214 (2005).
12. Chao, D. L., Halloran, M. E., Obenchain, V. J. & Longini, I. M. Jr. FluTE, a publicly available stochastic influenza epidemic simulation model. *PLoS Comput. Biol.* **6**(1): e1000656 (2010).
13. Merler, S. & Ajelli, M. The role of population heterogeneity and human mobility in the spread of pandemic influenza. *Proc. R. Soc. B.* **277**(1681), 557 – 565 (2009).
14. Epstein J. M., *et al.* Controlling Pandemic Flu: The Value of International Air Travel Restrictions. *PLoS ONE* **2(5)**, e401 (2007).
15. International migration. United Nation Statistics, Available at: http://unstats.un.org/unsd/demographic/sconcerns/migration/.
16. Keeling, M. J., Danon, L., Vernon, M. C. & Thomas, A. H. Individual identity and movement networks for disease metapopulations. *Proc. Natl. Acad.* Sci. USA **107,** 8866-8870 (2010).
17. Bajardi, P., Barrat, A., Natale, F., Savini, L. & Colizza, V. Dynamical Patterns of Cattle Trade Movements. *PLoS ONE* **6(5)**, e19869 (2011).
18. Gilsdorf, A., *et al.* Influenza A(H1N1)v in Germany: the first 10,000 cases. *Euro Surveill.* **14**, 34 (2009).
19. Schneeberger, A., *et al.* Scale-free networks and sexually transmitted diseases: A description of observed patterns of sexual contacts in Britain and Zimbabwe. *Sex. Trans. Dis.* **31**, 380 (2004).
20. Galvani, A. P. & May, R. M. Epidemiology – dimensions of superspreading. *Nature* **438**, 293-295 (2005).
21. Lloyd-Smith, J. O., Schreiber, S. J., Kopp, P. E. & Getz, W. M. Superspreading and the effect of individual variation on disease emergence. *Nature* **438,** 355-359 (2005).
22. Stehle. J., *et al.* Simulation of an SIR infectious disease model on the dynamic contact network of conference attendees. *BMC Med.* **9**, 87 (2011).
23. Barrat, A., Barthélemy, M., Pastor-Satorras, R. & Vespignani, A. The architecture of complex weighted networks. *Proc. Natl. Acad.* Sci. USA **101**, 3747-3752 (2004).
24. Chowell, G., Hyman, J. M., Eubank, S., & Castillo-Chavez, C. Scaling laws for the movement of people between locations in a large city. *Phys. Rev. E* **68**, 066102 (2003).

25. Pastor-Satorras, R. & Vespignani, A. Epidemic spreading in scale-free networks. *Phys. Rev. Lett.* **86**, 3200-3203 (2001).
26. Lloyd, A. L. & May, R. M. How viruses spread among computers and people. *Science* **292**, 1316-1317 (2001).
27. Colizza, V., Pastor-Satorras, R. & Vespignani, A. Reaction-diffusion processes and metapopulation models in heterogeneous networks. *Nature Phys*. **3**, 276-282 (2007).
28. Colizza, V., & A. Vespignani, A. Invasion threshold in heterogeneous metapopulation networks. *Phys. Rev. Lett.* **99**, 148701 (2007).
29. Meyers, L. A., Pourbohloul, B., Newman, M. E. J., Skowronskic, D. M. & Brunham. R. C.Network theory and SARS: predicting outbreak diversity. *Journal. Theor. Biol*. **232**, 71-81 (2005).
30. Balcan, D. & Vespignani, A. Phase transitions in contagion processes mediated by recurrent mobility patterns. *Nature Phys*. **7**, 581-586 (2011).
31. Rocha, L. E. C., Liljeros, F. & Holme, P. Simulated Epidemics in an Empirical Spatiotemporal Network of 50,185 Sexual Contacts *PLoS Comp. Biol*. **7**, e1001109 (2011).
32. Brockmann, D., Hufnagel, L. & Geisel, T. The scaling laws of human travel. *Nature* **439**, 462-465 (2006).
33. Colizza, V. & Vespignani, A. Epidemic modelling in metapopulation systems with heterogeneous coupling pattern: Theory and simulations. *J. Theor. Biol.* **251**, 450-467 (2008).
34. González, M. C., Hidalgo, C. A. & Barabási A. -L. Understanding individual human mobility patterns. *Nature* **453**, 779-782 (2008).
35. Song, C., Koren, T., Wang, P. & Barabasi A. -L. Modelling the scaling properties of human mobility. *Nature Phys.* **6**, 818 (2010).
36. Eurostat, Air transport measurement – passengers, Available at: http://epp.eurostat.ec.europa.eu/.
37. Viboud C., *et al.* Synchrony, waves, and spatial hierarchies in the spread of influenza. *Science*, **312** 447-451 (2006).
38. Decrop, A. & Snelders, D. Planning the summer vacation: an adaptable and opportunistic process. *Annals of Tourism Research* **31**(4), 1008 – 1030 (2004).
39. Gokovali, U., Bahar, O. & Kozak, M. Determinants of length of stay: a practical use of survival analysis. *Tourism Management* **28**, 736 – 746 (2007).
40. UK Office for National Statistics, Travel Trends 2007.

41. Eurostat, Tourism statistics at regional level, Available at: http://epp.eurostat.ec.europa.eu.
42. Lohmann, G., Albers, S., Koch, B. & Pavlovich, K. From hub to tourist destination – an explorative study of Singapore and Dubai's aviation-based transformation. *Journal of Air Transport Management* **15**, 205 – 211.
43. McKercher, B. & Lew, A. A. Distance decay and the impact of the effective tourism exclusion zones in international travel flows. *Journal of Travel Research* **42**, 159 (2003).
44. Rvachev & L. A., Longini, I. M. A model for the global spread of influenza. *Math. Biosci.* **75**, 3-22 (1985).
45. Sattenspiel, L. & Dietz, K. A structured epidemic model incorporating geographic mobility among regions. *Math. Biosci.* **128**, 71-91 (1995).
46. Keeling, M. J. & Rohani, P. Estimating spatial coupling in epidemiological systems: A mechanistic approach. *Ecol. Lett.* **5**, 20-29 (2002).
47. Anderson, R. M. & May, R. M. *Infectious Diseases of Humans: Dynamics and Control* (Oxford University Press, 1992).
48. Ball, F., Mollison, D. & Scalia-Tomba, G. Epidemics with two levels of mixing. *Ann. Appl. Probab.* **7**, 46-89 (1997).
49. Cross, P., Lloyd-Smith, J.O., Johnson, P.L.F. & Wayne, M.G. Duelling timescales of host movement and disease recovery determine invasion of disease in structured populations. *Ecol. Lett.* **8**, 587-595 (2005).
50. Cross, P., Johnson, P.L.F., Lloyd-Smith, J.O. & Wayne, M.G. Utility of $R_0$ as a predictor of disease invasion in structured populations. *J. R. Soc. Interface* **4**, 315-324 (2007).
51. Harris, T.E. *The Theory of Branching Processes* (Dover Publications, 1989).
52. Bailey, N. T. *The Mathematical Theory of Infectious Diseases*, 2$^{nd}$ ed. (Hodder Arnold, 1975).
53. Watts, D., Muhamad, R., Medina, D. C. & Dodds, P. S. Multiscale resurgent epidemics in a hierarchical metapopulation model. *Proc. Natl Acad. Sci. USA* **102**, 11157–11162 (2005).
54. Bajardi, P., Poletto, C., Ramasco, J.J., Tizzoni, M., Colizza, V. & Vespignani, A. Human Mobility Networks, Travel Restrictions, and the Global Spread of 2009 H1N1 Pandemic. *PLoS ONE* **6(1)**, e16591 (2011).
55. Catanzaro, M., Boguña, M. & Pastor-Satorras, R. Generation of uncorrelated random scale-free networks. *Phys. Rev. E* **71**, 027103 (2005).
56. Erdős, P. & Rényi, A. On random graphs. *Publ. Math*. **6**, 290 (1959)

57. Murray, J. D. *Mathematical Biology*, 3rd ed. (Springer, Berlin 2005).
58. Chu, Y. J. & Liu, T. H. On the shortest arborescence of a directed graph. *Science Sinica* **14**, 1396 (1965).

## Acknowledgments

The authors would like to thank Duygu Balcan and Alessandro Vespignani for fruitful interactions on the present work, and Hao Hu for discussions on the epidemic invasion tree. This work has been partially funded by the ERC Ideas contract n.ERC-2007-Stg204863 (EPIFOR) to VC, CP, and MT; the EC-ICT contract no. 231807 (EPIWORK) and the EC-FET contract no. 233847 (DYNANETS) to VC.

## Author contributions

Conceived and designed the experiments: CP, MT and VC. Performed the experiments: CP, MT and VC. Analysed the data: CP, MT and VC. All authors wrote, reviewed and approved the manuscript.

## Additional information

**Competing financial interests**

The authors declare no competing financial interests.

# Figures

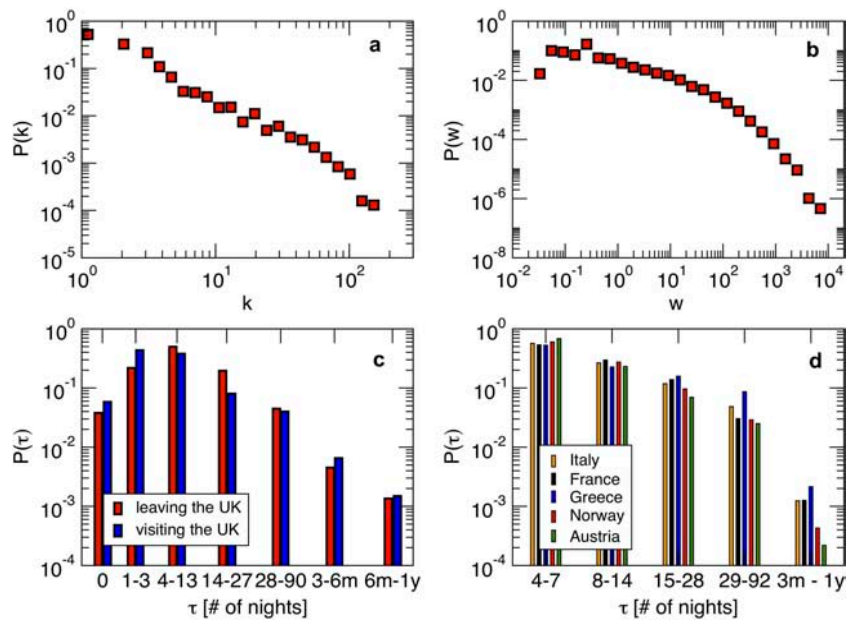

**Figure 1**. **Heterogeneous topology, traffic, and length of stay of the mobility network by air travel in Europe.** Statistical fluctuations are observed over a broad range of length and time scales. (**a**) The degree distribution $P(k)$ of the European airport network is characterized by large fluctuations indicating a heterogeneous topology of the system. (**b**) The distribution of weights in the European airport network is skewed and heavy-tailed. (**c**) The number of nights spent by foreign travellers visiting the UK (in blue) and by British travelling abroad (in red) spans several orders of magnitude, from 1 night to several months, considering all travel purposes and all countries of origin and destination. (**d**) The number of nights spent by tourists travelling to European countries on holiday is similarly broadly distributed. Here we show data for 5 selected countries of destination, considering only trips of 4 nights or more (source: Eurostat).

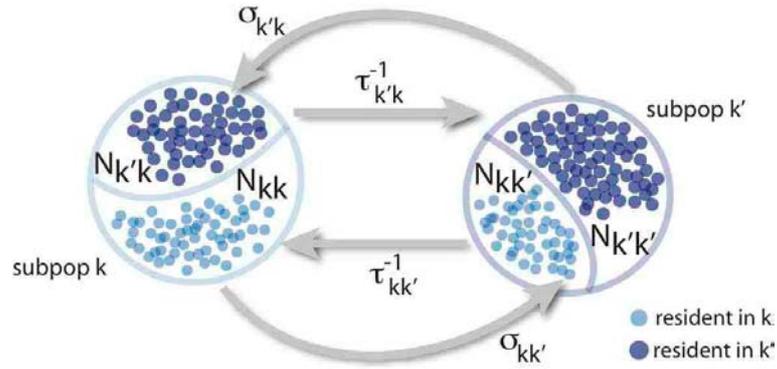

**Figure 2**. **Schematic representation of the non-Markovian travelling dynamics.** At any time the subpopulation with degree $k$ is occupied by a fraction of its own population $N_{kk}$ (individuals resident in the subpopulation with degree $k$ and currently located there) and a fraction of individuals $N_{k'k}$ whose origin is the neighbouring subpopulation with degree $k'$ and who are currently visiting the degree $k$ subpopulation. Travelling individuals from the subpopulation with degree $k$ (light blue particles) leave their home subpopulation to the subpopulation with degree $k'$ with rate $\sigma_{kk'}$ and return back with rate $\tau_{kk'}^{-1}$, where $\tau_{kk'}$ is the average time spent at destination. Here we assume that the length of stay is function of the destination only, namely $\tau_{kk'} \equiv \tau_{k'}$. This exchange of individuals between subpopulations is the origin of the transmission of the contagion process among the subpopulations.

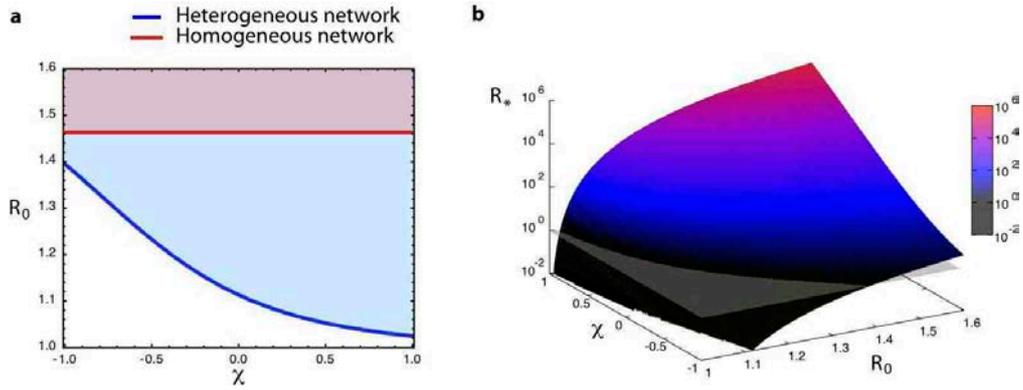

**Figure 3. Threshold condition for global invasion: analytical results.** (**a**) Invasion regions in the plane $(R_0, \chi)$ corresponding to the condition $R_* > 1$ of Eq. (5) for a heterogeneous metapopulation network (blue) with size $V = 10^4$ characterized by a power-law degree distribution $P(k) \propto k^{-\gamma}$ with $\gamma = 3$ and average degree $\langle k \rangle = 3$, and a homogeneous network (red) with same size and subpopulations degrees uniformly equal to $\langle k \rangle$. Other parameters are set equal in the two cases: diffusion rate $\sigma = 10^{-5}$ and scaling exponents $\theta = 1/2$ and $\phi = 3/4$ as in the worldwide airport network[9,26], average subpopulation size $\bar{N} = 10^3$, and assumed average length of stay $\bar{\tau} = 37$ to justify the timescale separation assumption and also allow the comparison with the numerical results. Since these parameters depend on the specific application under study, we report in the SI an exploration of additional parameter values. (**b**) Analytical surface of the global invasion threshold, $R_* = (R_0, \chi)$, as a function of the reproductive number $R_0$ and of the parameter $\chi$ tuning the heterogeneity of the length of stay. The surface is calculated in the case of the heterogeneous network considered in panel a.

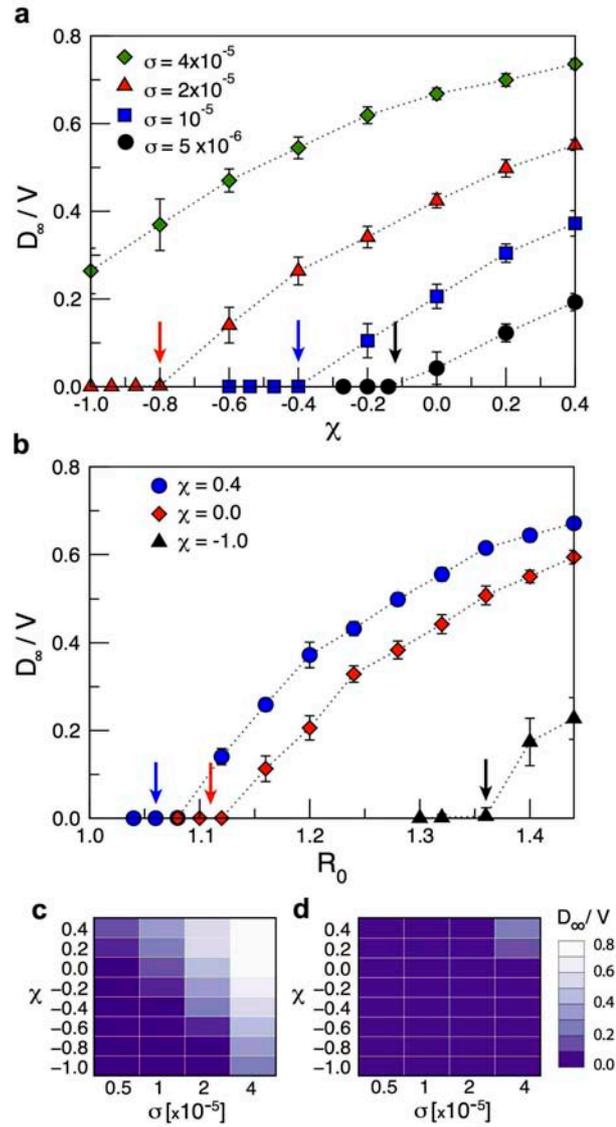

**Figure 4. Threshold condition for global invasion: simulation results.** (**a**) Average final fraction of infected subpopulations $D_\infty/V$ (symbols) and standard deviations (error bars) obtained from the simulations of an SIR epidemic with $R_0 = 1.2$ as a function of the exponent $\chi$ tuning the length of stay, for different values of the diffusion rate rescaling $\sigma$. Coloured arrows indicate the analytical value of the corresponding transition from no invasion to the global invasion regime. Error bars correspond to the standard deviation obtained from 500 stochastic simulations. (**b**) Average final fraction of infected subpopulations $D_\infty/V$ (symbols) and standard deviations (error bars) as a function the basic reproductive number $R_0$, for different values of the length of stay parameter $\chi$, and for diffusion rate rescaling $\sigma = 10^{-5}$. Coloured arrows as above. (**c,d**) Comparison of the simulated average fraction of infected subpopulations $D_\infty/V$ as a function of $\chi$ and $\sigma$ for a heterogeneous network (**c**) and a homogeneous network with same size

and average degree (**d**). Here $R_0 = 1.2$. In all panels the parameter values are kept equal to the ones used in Figure 3 to allow for the direct comparison between analytic and numerical results. In addition, the infectious rate is $\mu = 0.002$ (in arbitrary time units) to ensure the timescale approximation.

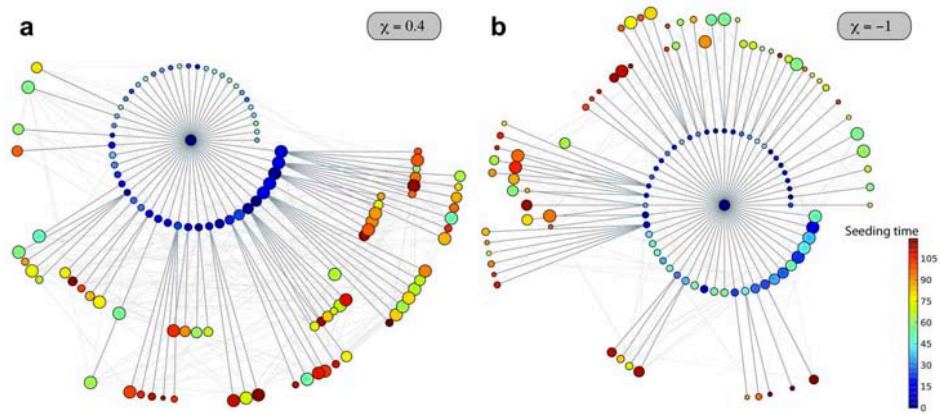

**Figure 5. Epidemic invasion trees.** The cases of a positively-correlated ($\chi = 0.4$) and negatively-correlated ($\chi = -1$) length of stay are shown. The synthetic network is characterized by a power-law distribution $P(k) \propto k^{-\gamma}$ with $\gamma = 2.1$, size $V = 10^4$ subpopulations and average population size $\overline{N} = 10^3$. An SIR dynamics starting from the same seeding node (at the centre of each visualization) is simulated, with $R_0 = 1.8$, $\sigma = 10^{-4}$, $\bar{\tau} = 37$, and $\mu = 0.002$ (in arbitrary time units) to ensure the timescale approximation. Only the first 120 nodes to be infected are displayed for the sake of visualization, on successive layers of invasion. Larger width grey links correspond to the paths of infection and lighter grey ones to the existing connections among visible nodes. Nodes are colour coded according to the time of their seeding, and their size scales with their degree; nodes in the first layer are ordered according to their degree to highlight the role of different degree nodes in the hierarchical invasion pattern in the two cases.

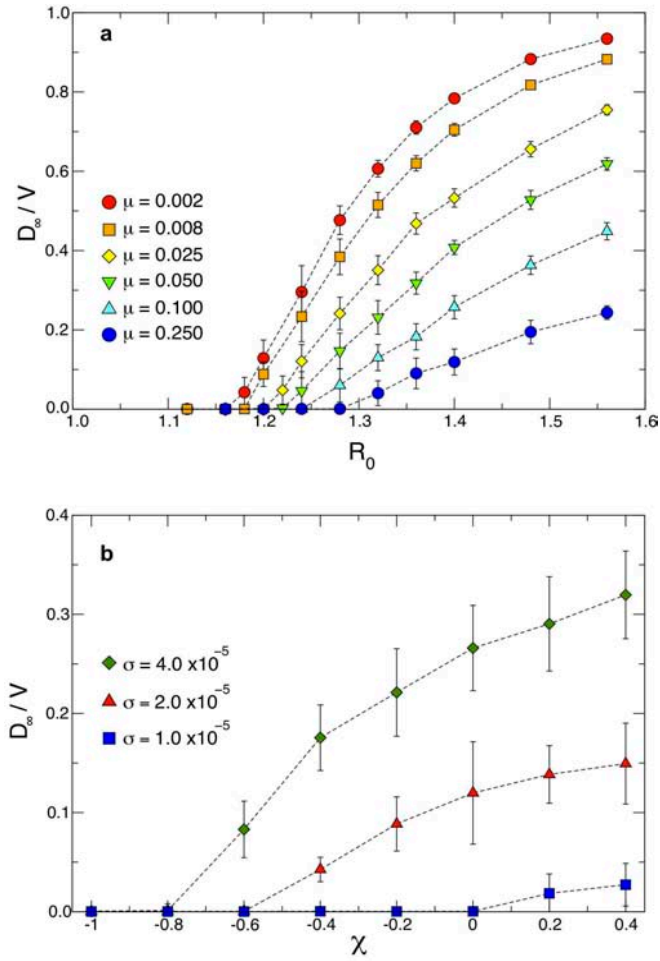

**Figure 6. Threshold condition for global invasion for shorter infectious periods: simulation results.** (**a**) Average final fraction of infected subpopulations $D_\infty/V$ (symbols) and standard deviations (error bars) obtained from simulations as a function of the basic reproductive number $R_0$, for different SIR-like disease characterized by different infectious periods. Here we consider a diffusion rate rescaling $\sigma = 4 \cdot 10^{-5}$ and $\chi = -1$. All other parameters are set as in Figure 4b. (**b**) Average final fraction of infected subpopulations $D_\infty/V$ (symbols) and standard deviations (error bars) obtained from simulations as a function of the length of stay parameter $\chi$, for different values of the diffusion rate rescaling $\sigma$. Here the same parameters of Figure 4a are used, except for a shorter infectious period, corresponding to $\mu = 0.05$.

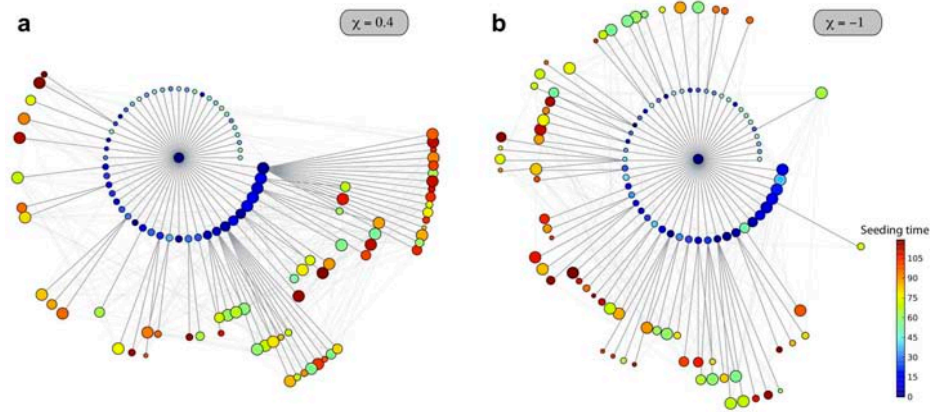

**Figure 7. Epidemic invasion trees for shorter infectious periods.** The cases of a positively-correlated ($\chi = 0.4$) and negatively-correlated ($\chi = -1$) length of stay are shown. The synthetic network is characterized by a power-law distribution $P(k) \propto k^{-\gamma}$ with $\gamma = 2.1$, size $V = 10^4$ subpopulations and average population size $\bar{N} = 10^3$. An SIR dynamics starting from the same seeding node (at the center of each visualization) is simulated, with $R_0 = 1.8$, $\sigma = 10^{-4}$, $\bar{\tau} = 37$. All parameters are thus set equal to the ones in Figure 5, except for the shorter value of the infectious period considered, corresponding to $\mu = 0.05$. As in Figure 5, only the first 120 nodes to be infected are displayed for the sake of visualization, on successive layers of invasion. Larger width grey links correspond to the paths of infection and lighter grey ones to the existing connections among visible nodes. Nodes are colour coded according to the time of their seeding, and their size scales with their degree; nodes in the first layer are ordered according to their degree to highlight the role of different degree nodes in the hierarchical invasion pattern in the two cases.

# Tables

**Table 1. Definition of the degree-block variables.**

| Degree-block variable | Definition |
|---|---|
| $k$ | Degree, i.e. the number of connections of a subpopulation |
| $V_k$ | Number of subpopulations with degree $k$ |
| $N_k$ | Individuals resident in subpopulations of degree $k$ |
| $N_{kk'}$ | Individuals resident in subpopulations of degree $k$ located in subpopulation of degree $k'$ |
| $\tau_k$ | Length of stay of individuals at subpopulations with degree $k$ |
| $w_{kk'}$ | Number of individuals leaving a subpopulation of degree $k$ to a subpopulation of degree $k'$ |
| $\sigma_{kk'}$ | Leaving rate of travellers from subpopulations of degree $k$ to subpopulations of degree $k'$ |
| $\sigma_k$ | Total leaving rate out from subpopulations of degree $k$ |
| $T_{kk'}$ | Total volume of people travelling on a connection between subpopulations of degree $k$ and subpopulations of degree $k'$ |